%% file: main.tex
\documentclass[12pt, USletter]{article} 

\usepackage[papersize={7in,9.25in}]{geometry} 
\RequirePackage{fancyhdr} 
\usepackage{natbib} 
\usepackage{times} 

\usepackage{setspace} 
\usepackage{amsmath}
\usepackage{amssymb}
\usepackage{booktabs}
\usepackage{multirow}
\usepackage{verbatim}
\usepackage{enumitem}
\usepackage{bbm}
\usepackage{hyperref}
\usepackage{subcaption}
\usepackage{adjustbox}
\usepackage{graphicx}
\usepackage{booktabs}
\usepackage{multirow}
\usepackage{array,booktabs}
\usepackage{hhline}
\usepackage{algorithm2e}
\usepackage[noend]{algpseudocode}
\usepackage{wrapfig}
\usepackage{etoolbox}
\usepackage{setspace}
\usepackage{longtable}
\usepackage{rotating}
\usepackage{titlesec}
\usepackage{xcolor}
\usepackage{titling}
\usepackage{amsmath, amssymb, amsfonts}  
\usepackage{relsize}
\usepackage[normalem]{ulem}
\usepackage{graphicx}
\usepackage{tabularx}
\usepackage{booktabs}
\usepackage{url}
\useunder{\uline}{\ul}{}

\newcolumntype{L}[1]{>{\raggedright\arraybackslash}p{#1}}
\usepackage[toc,page]{appendix}
\usepackage{threeparttable}
\usepackage{pdflscape}
\usepackage{secdot}
\sectiondot{subsection}
\usepackage[labelfont=bf]{caption}

\titlespacing{\section}{2pt}{2pt}{2pt}
\titlespacing\subsection{2pt}{2pt}{2pt}
\titlespacing\subsubsection{2pt}{2pt}{2pt}
\AtBeginDocument{
\setlength\abovedisplayskip{2pt}
\setlength\belowdisplayskip{2pt}}

\titleformat{\paragraph}
{\normalfont\small\itshape}{\theparagraph}{1em}{}
\titlespacing*{\paragraph}{0pt}{3.25ex plus 1ex minus .2ex}{1.5ex plus .2ex}

\usepackage{acronym}
\newcommand{\shorttitle}[1]{\gdef\@shorttitle{#1}}
\shorttitle{\fontsize{9}{10}\selectfont When KD meets LTH}

\graphicspath{ {./images/} }

\fancypagestyle{plain}{
\fancyhf{}
\fancyfoot[L]{\fontsize{9}{10}\selectfont\textit{}\hspace{5mm}\hfill\textbf{\thepage}}
}

\usepackage{titlesec}
\titlespacing*{\section}{0pt}{0.1\baselineskip}{0.2\baselineskip}

\title{\fontsize{20}{17}\selectfont \textbf{Enhancing Scalability in Recommender Systems through Lottery Ticket Hypothesis and Knowledge Distillation-based Neural Network Pruning} \\
\vspace{0.1in}\fontsize{12}{15}\selectfont \textit{Regular Paper}
    \vspace{-2.2em}}

\author{Rajaram R, IIT Madras, India, raajaram100@gmail.com\\
 Manoj Bharadhwaj, IIT Madras, India, cs20s056@cse.iitm.ac.in\\
 Vasan VS, Pravartak, IIT Madras, India, vasan.vs@gmail.com\\
 Nargis Pervin*, IIT Madras, India, nargisp@iitm.ac.in}
\vspace{3ex}
\date{}



\newcounter{example}[section]
\titleformat{\section}
{\fontsize{13}{15}\bfseries}{\thesection}{1em}{}

\titleformat{\subsection}
{\fontsize{12}{15}\bfseries\itshape}{\thesubsection}{1em}{}

\titleformat{\subsubsection}
{\fontsize{12}{15}\bfseries}{\thesubsubsection}{1em}{}

\begin{document}

\maketitle
\vspace{-1.9cm}
\input{P_0_Abstract}
\input{P_0_1_Key_Words}
\input{P_1_Introduction}

\input{P_2_Related_Works}
\input{P_3_Solution_Details}
\input{P_4_Dataset_Desc}
\input{P_5_Experimental_Results}
\input{P_7_Conclusion}
\singlespacing
\bibliographystyle{misq}
\bibliography{references}

\end{document}

%% file: P_0_Abstract.tex
\section*{Abstract}


This study introduces an innovative approach aimed at the efficient pruning of neural networks, with a particular focus on their deployment on edge devices. Our method involves the integration of the Lottery Ticket Hypothesis (LTH) with the Knowledge Distillation (KD) framework, resulting in the formulation of three distinct pruning models. These models have been developed to address scalability issue in recommender systems, whereby the complexities of deep learning models have hindered their practical deployment. With judicious application of the pruning techniques, we effectively curtail the power consumption and model dimensions without compromising on accuracy. Empirical evaluation has been performed using two real world datasets from diverse domains against two baselines. Gratifyingly, our approaches yielded a GPU computation-power reduction of up to 66.67\%. Notably, our study contributes to the field of recommendation system by pioneering the application of LTH and KD.

%% file: P_0_1_Key_Words.tex

%% file: P_1_Introduction.tex
\section*{Introduction}

Deep Learning has become a transformational force in the ever-expanding arena of artificial intelligence, revolutionizing various fields with its outstanding capacity to analyze massive volumes of data and derive insightful knowledge~\citep{article31}. Specifically, Convolutional Neural Networks (CNNs) and Recurrent Neural Networks (RNNs) are used to solve various tasks across domains including computer vision, natural language processing, and recommender systems. This is partly due to its illustrious performance and the allure of learning fine-grained feature representations.
Traditionally, collaborative filtering based recommendation systems \citep{10.1145/963770.963772} use prior user choices or activities to suggest new products or items to match the user preferences. For example, in movie recommendation system, user's historical data on viewing and browsing along with movie features (like actor, director, genre, etc.) could be used to generate recommendations. However, these traditional methods suffer from well-known data-sparsity and scalability issues.


While prior works have implemented deep Learning based recommender systems and focused mainly on improving accuracy~\citep{7124857, DBLP:journals/corr/ZhangYS17aa}, scalability issues have largely been ignored~\citep{1460447}. 
Specifically, the complexity of deep learning based techniques increases with the number of hyperparameters and the number of layers (size) in the model. The task becomes daunting in the practical scenario, when the model has to be scaled across multiple platforms and interfaces, from mobile phones to complex GPU servers. One plausible way to address this scalability issue is to make the model smaller, (i.e., lowering complexity) so that the latency, computation time, and power decreases~\citep{Kumar2018RecommendationST}, and thereby the models can be deployed on any platform. While the potential of CNN in recommendations is well accepted in the research community, advances in this direction is still in its nascent stage~\citep{1206588}. 
In order to address this issue, one intriguing concept is the Lottery Ticket Hypothesis, which draws an analogy to solving puzzles. Imagine a hard issue is represented by a challenging puzzle (complex neural network). The Lottery Ticket Hypothesis postulates that inside a complex puzzle (neural network), there may be smaller, vital components that contribute to its overall effectiveness, just as some smaller pieces of the puzzle (smaller neural networks) may hold the secret to its solution. These ``winning tickets" can be found and used to increase the complex puzzle's (neural network's) effectiveness and efficiency, which will hasten the learning process. Another significant concept is of Knowledge distillation, which resembles the relationship between a teacher and student, is another important concept. In this situation, a teacher, symbolizing a big, complicated model, teaches a student, representing a tiny, simple model. Knowledge distillation includes the teacher model communicating its insights to the student model in a distilled and more understandable form, much like how a teacher might reduce complex subjects down into simpler lessons.

This paper proposes a novel method to prune a neural network efficiently which can be used to deploy on edge devices. To achieve this objective, we propose three pruning models that combine the Lottery ticket hypothesis \citep{frankle2019lottery} with Knowledge Distillation framework \citep{hinton2015distilling}. Our first model, namely, Structured Pruning with Show Attend and Distill (SP-SAD)  aims to use structured pruning to make the model as sparse as possible during the training itself. In the second model, Lottery Ticket Hypothesis with Show Attend and Distill (LTH-SAD), we use the Lottery Ticket Hypothesis to re-initialize the parameters of the student model after pruning with the ones that were there during the start. In the third model, Sparse Student with Show Attend and Distill (SS-SAD), here the student model is first pruned using Lottery Ticket Hypothesis, this pruned model now acts as a student model in knowledge distillation process. In this context, user devices capable of conducting computations locally, as opposed to relying on a centralized server. Employing this pruning technique results in quicker response time and faster computation. Notably, for edge devices (eg., user's handheld mobile devices) optimizing power consumption is of utmost importance, which is attainable through the well-designed curation of a pruned model. 

To evaluate the efficacy of our proposed approach (SP-SAD, LTH-SAD, SS-SAD), we have performed extensive experiments using two public datasets, where CIFAR-100 is from computer vision domain used for object detection, and the other is movie datasets, (IMDb and TMDb), used for recommendations. The proposed approaches (SP-SAD, LTH-SAD, SS-SAD) have been compared with two baselines. We have used the power consumption and the model size as the metric to compare the scalability of the models. According to a prior study by \cite{Kumar2018RecommendationST}, with the reduction in the model size, the power and training time decreases. The experimental findings reveal that our proposed methods achieve comparable accuracy (up to 32.08\% , 25.10\% improvement in MSE and MAE, respectively) with reduced GPU power consumption (up to 66.67\%) with 45\% reduction in model size which in turn reduces the carbon footprint. The primary contributions of our study are the following:
\begin{enumerate}
    \item Improving structured pruning during training of a convolutional neural network using an attention-based Knowledge Distillation technique. 
   
    \item Addressing the scalability issue of recommender systems with reduction in power consumption without compromising on accuracy.
\end{enumerate}

To the best of our knowledge, this work is the first attempt to apply the lottery ticket hypothesis combined in the knowledge distillation process to train a recommender system architecture.

%% file: P_2_Related_Works.tex
\section*{Related works}

\subsubsection*{CNN in Recommendation systems}
Initially designed for image processing tasks, convolutional Neural Networks (CNNs) have gained attention in recommendation systems due to their ability to capture complex patterns and learn hierarchical representations from data. Several studies have explored the integration of CNNs into recommendation systems to enhance their performance and address challenges such as cold start problems and sparsity issue. For instance, CNNs incorporated document context information into recommendation models, improving performance in document recommendation tasks \citep{8716661}. Similarly, \cite{zhou2018deep} employed CNNs to model user interests and interactions for personalized click-through rate predictions in e-commerce platforms. 
However, performance of CNNs rely on the intensive parameter list and a large training dataset. On the other hand, traditional machine learning techniques, content-based filtering, and small neural networks are under-parameterized and run faster \citep{gordon2018morphnet}. 
Thus, a plausible way to speed up recommendation systems is to use CNNs and make them sparser by pruning a neural network \citep{9244787}. Typically, there are two pruning techniques for neural networks, structured and unstructured pruning \citep{molchanov2017pruning}. 
Structured pruning involves removing entire structures or groups of parameters from a neural network. These structures can include whole filters, channels, or even layers \citep{9244787}. For example, if a layer contains 64 filters, structured pruning may remove a fixed percentage (e.g., 30\%) of filters, resulting in a reduced layer with 45 filters \citep{xia2022structured}.
Unstructured pruning involves removing individual parameters (weights) from a neural network, regardless of their position in the model \citep{kdpri}.

\subsubsection*{Lottery Ticket Hypothesis}
``Lottery Ticket Hypothesis" \citep{frankle2019lottery} is an intriguing pruning technique that casts doubt on how we understand deep learning models and their underlying structure. The theory states that hidden sub-networks, referred to as ``winning tickets," inside an originally over-parameterized network may achieve comparable or even better performance with much fewer parameters. Applications of the lottery ticket hypothesis are majorly on object detection and various computer vision tasks. \cite{Girish_2021_CVPR} have mentioned the application of the Lottery Ticket Hypothesis on embedded edge devices.

\subsubsection*{Knowledge Distillation}
Knowledge distillation \citep{hinton2015distilling} transfers the ``knowledge" in a neural network by moving the information from a large ``teacher" network to a smaller ``student" network. Knowledge distillation opens up new possibilities for deployment in resource-constrained contexts by encapsulating the essence of the teacher's expertise while maintaining performance. Extending Knowledge Distillation, a few recent studies have shown significant improvement in model compression \citep{romero2015fitnets, 8100237,tian2022contrastive,9857422}. This approach recently has as well found ample applications in transfer learning in cross-domain \citep{orbesarteaga2019knowledge,7953145} and continual learning \citep{li2017learning,10.1007/978-3-030-01219-9_27}.

In this paper, we propose three models based on pruning technique that aims to reduce power consumption with an improvement in prediction accuracy measured by MSE and MAE. 

%% file: P_3_Solution_Details.tex
\section*{Methodology}
\subsubsection*{Methodology Overview}
This study focuses on developing a novel scalable pruning framework, where the goal is to reduce the power consumption of recommender systems. Overall, a smaller student model is trained using the knowledge distilled from a larger teacher model, and over the period, pruning of the student model is performed to make it sparse. 
\begin{figure}[ht]
    \centering
    \includegraphics[width=0.7\textwidth, height= 6.5cm]{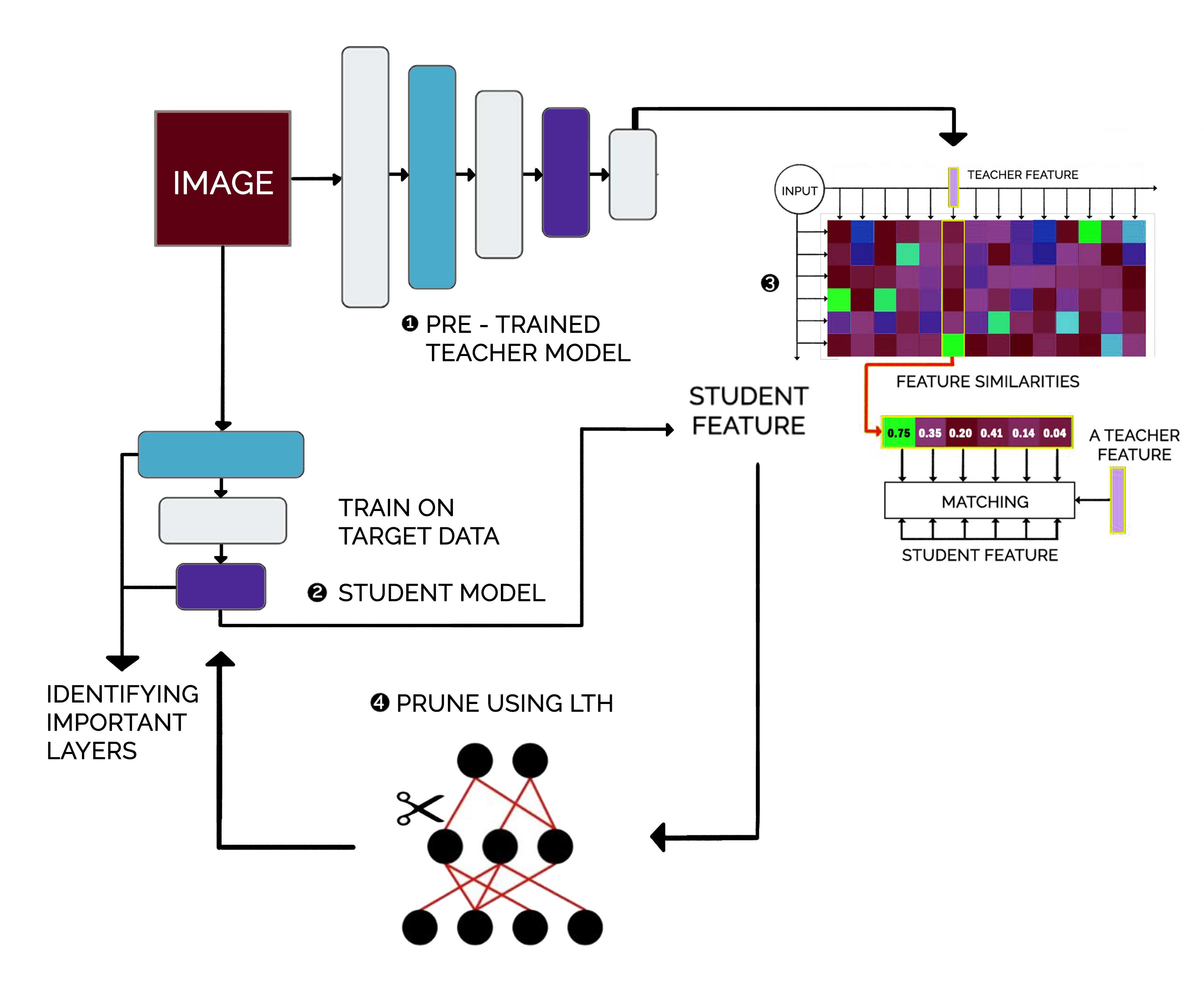}
    \caption{Pruning Framework using Knowledge Distillation and Lottery Ticket Hypothesis}
    \label{fig:Schematic_diag}
\end{figure}
Figure~\ref{fig:Schematic_diag} shows the framework for pruning using KD and LTH. First, the input image is fed to both the teacher and the student model; the features of the image has been extracted from each layer of the teacher model as well as from the student model. Now each layer of the teacher will transfer its features to the suitable student layer through attention mechanism. After this whole process, training of the student model begins. During the back-propagation while training, dynamic structured pruning is done to make the model more sparse. And this is repeated till the "Winning Ticket" is found, i.e., till a small sparse neural network which yields nearly equivalent or greater accuracy than the teacher model is attained. Based on the schematic diagram \ref{fig:Schematic_diag}, here is a detailed overview of how our models are structured:
\begin{enumerate}
    \item Structured Pruning with Show Attend and Distill (SP-SAD): This approach uses structured pruning to make the student model as sparse as possible during the training itself. Here, in structured pruning we re-initialize the parameters of the student model after pruning with the ones of obtained in the previous step.
    \item Lottery Ticket Hypothesis with Show Attend and Distill (LTH-SAD): In this approach, we use the Lottery Ticket Hypothesis to re-initialize the parameters of the student model after pruning with the ones that were there during the start.
    \item  Sparse Student with Show Attend and Distill (SS-SAD): Here, the student model is first pruned using Lottery Ticket Hypothesis (LTH), and then the training happens. Here, the output of LTH acts as a student model in the Knowledge distillation process.
\end{enumerate}

\subsubsection*{Methodology Details}
 Show attend, and Distill \citep{ji2021show} is a technique to distill the weights of a neural network where
\begin{itemize}
    \item Let $F^T=\{F_{1}^T,\dots, F_{n}^{T}\}$ be the features representing the outputs of various layers of the neural network where $F_i^T$ represents features of $i^{th}$ layer and 
    $F_{n}^{T}$ represents the features of the pre-final layers of the teacher model, $n+1$ being the number of layers. 
    \item Similarly, let $ F^S=\{F_{1}^S, ..., F_{m}^{S}\}$ be the feature set of the student model where $F_i^S$ represents features of $i^{th}$ layer and $F_{m}^{T}$ represents the features of the pre-final layers of the teacher model, $m+1$ being the number of layers.
\end{itemize}
The next step is knowledge transfer which involves global average pooling of features from both the teacher and student models. The pairing intensity for knowledge transfer is determined by the distance computed from channel-wise pooled features. To calculate feature similarities, the method employs the concept of query and key pairs from attention mechanisms. Queries \(Q_{T}\) are generated from teacher features, and keys \(K_{S}\) from student features. They are represented as follows:
\begin{align*}
    Q_{T} = \emph{f}_{\emph{Q}}(W_{n}^\emph{Q}.\phi^{hw}(F_{n}^\emph{T}))~~~~~ \text{and}~~~~
    K_{S} = \emph{f}_{\emph{K}}(W_{m}^\emph{K}.\phi^{hw}(F_{m}^\emph{S}))
\end{align*}
Here $\phi^{hw}(.)$ represents a global average pooling, and $\emph{f}_{\emph{Q}}$, $\emph{f}_{\emph{K}}$ are the activation functions of the query and the key. $(W_{n}^\emph{Q}$ and $(W_{m}^\emph{K}$ are the linear transition parameters for $n^{\text{th}}$ key and $m^{\text{th}}$ query.
Using softmax, probabilities are calculated with these keys and queries:
\begin{equation}
    \alpha_{t} = \text{Softmax}([Q_{n}^\emph{T}W_{1}^\emph{Q-K}K_{t,1} + ((P_{n}^\emph{T})^{\intercal})P_{1}^\emph{S}/\sqrt{d}, ..., Q_{n}^\emph{T}W_{m}^\emph{Q-K}K_{t,m} + ((P_{n}^\emph{T})^{\intercal})P_{m}^\emph{S}/\sqrt{d}])
\end{equation}
Here, \(W_{S}^\emph{Q-K}\) are the bi-linear weights, and \(P_{m}^\emph{S}\), \(P_{n}^\emph{T}\) are positional encodings. \(\alpha_{t}\) captures the probability of transferring knowledge from the teacher to the student.
The distillation term is represented as:
\begin{equation}
    \mathcal{L}_{Attention} = \sum_{t}\sum_{S}\alpha_{n,m} \parallel \widetilde{\phi}^{C}(h_{n}^T)-\widetilde{\phi}^{C} (\hat{h}_{m}^{S}) \parallel_{2}
\end{equation}
Where \(\widetilde{\phi}^{C}\) combines channel average pooling with $L_2$ norm. The loss function for knowledge distillation is given by:
\begin{equation}
    \mathcal{L}_{Student} = \mathcal{L}_{class} + \beta\mathcal{L}_{Attention}
\end{equation}
Here, \(\mathcal{L}_{class}\) is the classification loss, and \(\beta\) balances the attention loss.
This above equation shows the loss function for the knowledge distillation, where $\mathcal{L}_{class}$ is the ground truth classification loss which is calculated by cross-entropy loss. 
During the backprobagation this loss function is used to train the student model.

\subsubsection*{Algorithm}
Our work aims to increase the sparsity of a neural network iteratively without a significant drop in accuracy. To do this, we use the concept of attention-guided knowledge distillation \citep{xu2016show} and a structured pruning technique. We use a teacher model with $n+1$ layers and a student model with $m+1$ layers and $m \ll n$. The algorithmic steps have been discussed in Algorithm~\ref{algorithm}.

\begin{algorithm}[H]
\SetAlgoLined
\SetKwInOut{Input}{Input}
\SetKwInOut{Output}{Output}
\Input{Pre-trained Teacher Network with $n$ layers}
\Output{A fully trained sparse student model with m layers}


\textbf{Training Loop}:
\For{each epoch}{
Compute $F^T$ and $F^S$ \\
$L$ = Show Attend and Distill$(F^T, F^S)$\;
Back-propagation of the student model and optimization of $L$\;
\uIf{epoch \% args.prune\_every == 0}{
    Extract and prune the weight mask of the student model\;
    Reinitialize the student model\;   
  }
}

\caption{Iterative Pruning with Show Attend and Distill}
\label{algorithm}
\end{algorithm}
\vspace{2pt}
Here, $F^T$ and $F^S$ are the feature sets of the teacher and student model, and $L$ is the loss obtained from the Show Attend and Distill model. To reinitialize the model, the same weights that were used in the previous step is used in case of structured pruning and in case of LTH-SAD, the weights are reinitialized using the lottery ticket hypothesis.


%% file: P_4_Dataset_Desc.tex
\section*{Dataset Description}
\subsubsection*{CIFAR100 dataset}
CIFAR-100 \citep{articlecifar} dataset consists of 60,000 color photos which are 32 × 32 sized color images for 100 object classes (with 600 photos in each class). 
We split the dataset into 50K training and 10K validation images. Each class represents a distinct category of object, including cars, birds, dogs, cats, etc. The horizontal flipping and random cropping are used for data augmentation.

\subsubsection*{Movie Dataset}
 We used two movie datasets IMDb and TMDb obtained from \cite{article31}. Based on the title and year of the film's release, these datasets were aggregated into a single dataset containing 4317 films released between 1916 and 2016. A training dataset with movies released between 2000 and 2013 (3819 movies) and a test dataset with movies released after 2013 (498 movies) were created from the dataset. 
 Pre-processing procedures like data cleansing, data transformation, and feature extraction are carried out on both training and test data. The features are divided into four groups: social media, textual, categorical, and numerical features as listed in Table~\ref{tab:featureset}.
\begin{table}[htbp]
    \centering
    \small
    \begin{tabular}{llll}
        \toprule
        \textbf{Features} & \textbf{Description} & \textbf{Features} & \textbf{Description} \\
        \midrule
        \textbf{Numerical features} & Budget Cost & \textbf{Textual Features} & Movie Title \\
        & Time Duration & & Plot Keywords \\
        & Total Companies & & Overview \\
        & Release Day & & Tagline \\
        & Release Month & & \\
        & Release Year & & \\
        & Total Languages & & \\
        \textbf{Social Features} & Actor/Actress Likes (FBL) & & Director Likes \\
        & Cast Likes & & Crew FBL \\
        \textbf{Categorical Features} & Production Countries & & Genres \\
        & Content Rating & & \\
        \bottomrule
    \end{tabular}
    \caption{Categorised Features}
    \label{tab:featureset}
\end{table}


%% file: P_5_Experimental_Results.tex
\section*{Experiments Findings}
In this section, we present the experimental findings to assess the performance of the three proposed approaches, namely, SP-SAD, LTH-SAD, SS-SAD algorithms. All the experiments were carried out in Python 3.10.9 and having Linux-X11 (Ubuntu 22.04.2 LTS) computer specifications as CPU: 8Core/16 threads, RAM:64GB, NVIDIA Corporation GP102 [GeForce GTX 1080 Ti]. To verify that the results are not dataset dependant, we have chosen datsets from two distinct domains with diverse application scenarios.

\subsubsection*{Evaluation Metrics}
To quantify the performance of our model, we employ standard metrics such as accuracy, MAE, and MSE, and  compute GPU power consumption to measure scalability of the system.

\subsubsection*{Accuracy}
Accuracy measures the ratio of correct predictions to the total number of predictions. 
\[
\text{Accuracy} = \frac{\text{Number of Correct Predictions}}{\text{Total Number of Predictions}}
\]
\subsubsection*{Mean Absolute Error (MAE) and Mean Squared Error (MSE)}
Mean Absolute Error (MAE) calculates the average absolute difference between the predicted values and the actual values. Mean Squared Error (MSE) on the other hand measures the average of the squared differences between the predicted values and the actual values.

$\text{MAE} = \frac{1}{n} \sum_{i=1}^{n} |y_{i} - \hat{y}_{i}|$ ~~~~~and ~~~~$\text{MSE} = \frac{1}{n} \sum_{i=1}^{n} (y_{i} - \hat{y}_{i})^2$ \\
where \(y_{i}\), \(\hat{y}_{i}\) are the actual and predicted values of the target variable (ground truth) of $i^{th}$ observation, respectively.

\subsubsection*{Experimental Findings on CIFAR100 Dataset}
The effectiveness of the suggested approaches in terms of accuracy, has been demonstrated by comparison with SAD (Show Attend \& Distill) \citep{ji2021show}
   that incorporates the attention mechanism in the knowledge distillation mechanism.    
\begin{table}
    \small
    \centering
    \captionsetup{justification=centering} 
    \begin{tabular}{ |p{2.3cm}|p{6cm}|p{1cm}|p{1.8cm}|p{1cm}|p{2cm}| }
     \hline
     \textbf{Parameters} & \textbf{Description} & \textbf{SP-SAD} & \textbf{LTH-SAD} & \textbf{SS-SAD} & \textbf{SAD} \\
     \hline
     Total Epochs & \#training iterations & 160 & 1200 & 200&240 \\
     \hline
     Pruning Rate & \% of weights removed & 10\%& 5\%& 5\%&- \\
     \hline
     Learning Rate & Step size to update model parameters & 0.05&0.05 &0.0005 & 0.05\\
     \hline
     Prune every & Interval duration (in epochs) till the model is pruned & 20& 75& 20& -\\
     \hline
     Learning Decay & rate to lower the learning rate &0.001 &0.0005 & -& 0.05\\
     \hline
     Learning Rate Scheduling & When the learning rate is altered (in epochs) & 60,120&170,340,510 & -&150,180,210 \\
     \hline
     Beta $(\beta)$ & Controls the overall attention loss & 100& 50& 200& 200\\
     \hline
     Temperature (T) & normalizes the softmax values & 4&4 &4 & 4\\
     \hline
     Alpha $(\alpha)$ & Controls the teacher attention loss &0.9 &0.8 &0.9 &0.9 \\
     \hline
     Batch Size & Number of images fed in a batch & 64&64 &64 &64 \\
     \hline
     \end{tabular}
     \caption{Hyperparameter Tuning for SP-SAD, LTH-SAD, SS-SAD, and SAD}
     \label{HyperparameterTable}
\end{table}
Here all the mentioned models were trained using WRN-40-2 \citep{articlewrn} as their teacher model and WRN-16-2 \citep{articlewrn} as their student model. The optimal hyperparameters used for the models are presented in Table~\ref{HyperparameterTable}. Table~\ref{tab: CIFAR100} depicts the comparison of our three proposed approaches, SP-SAD, LTH-SAD, SS-SAD with SAD. It should be noted that while the baseline contains all the parameters (100\%), we apply pruning on the three proposed methods by varying the pruning percentage from 0\%-100\% with a pruning rate of 5\% -10\%. Finally, the wining ticket is chosen based on the best accuracy obtained. From Table~\ref{tab: CIFAR100} it can be observed that all the three proposed methods, SP-SAD, LTH-SAD, and SS-SAD achieve comparable accuracy (73\% (60\% pruned), 73.03\% (60\% pruned), 73.71\% (70\% pruned), respectively for SP-SAD, LTH-SAD, and SS-SAD) with a reduced model size in comparison to SAD. 
\begin{table}
    \centering
    \captionsetup{justification=centering}
    \begin{tabular}{ |p{3cm}|p{2cm}|p{3cm}|  }
     \hline
     Method & Accuracy & \% Size reduction \\
     \hline
     SAD & 75.47\% & 0\% \\
     SP-SAD & 73.00\% & 60\% \\
     LTH-SAD & 73.03\% & 60\% \\
     SS-SAD & 73.71\% & 70\% \\
     \hline
    \end{tabular}
    \caption{Comparison with CIFAR100 Dataset}
    \label{tab: CIFAR100}
\end{table}

\subsubsection*{Experimental Findings on Movies Dataset}
We experimented our models with a different dataset from movie domain and compared with SAD and DCNN \cite{article31}. DCNN ~\citep{article31} discusses a Convolutional Neural Network (CNN) based approach to predict movie ratings based on movie attributes. We performed the experiments on the four types of features (numerical, social, categorical, and textual) independently. 
 Table~\ref{Table4} depicts the performance comparison of the three models against MSE and MAE. The boldface denotes the best performing value while the second best value is underlined. It is evident from the table that LTH-SAD strategy achieves the lowest MSE and MAE with social and textual features. LTH-SAD with social features accounts for MAE and MSE values as 0.81 and 1.106 with an improvement of up to 7.9\% and 10\%, respectively. Similarly, LTH-SAD with textual features achieves 25.1\% and 32\% improvements on MAE (0.871) and MSE (1.319) values. This could be attributed to the fact that it most likely combines sophisticated information transferred with attention mechanisms. 
However, the observed performance of LTH-SAD with numerical features diverges from its anticipated outcome and require further exploration. 
\begin{figure}[ht]
    \centering
    \begin{subfigure}{0.45\textwidth}
        \centering
        \includegraphics[width=1.2\linewidth]{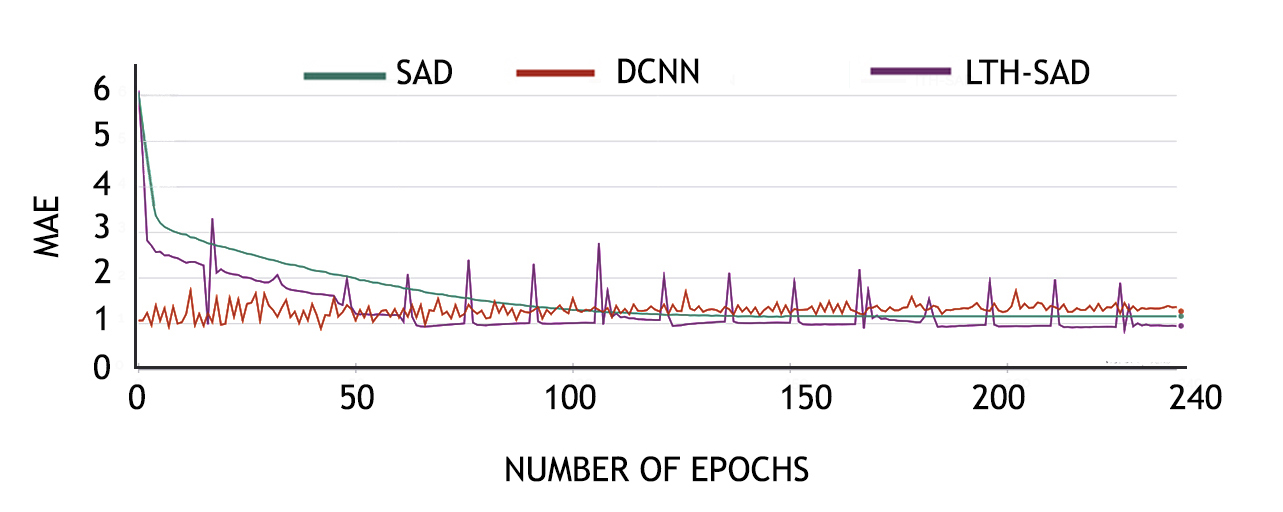}
        \caption{Comparison of MAE of LTH-SAD, SAD \& DCNN}
        \label{fig:topical_mae}
    \end{subfigure}
    \hfill
    \begin{subfigure}{0.45\textwidth}
        \centering
        \includegraphics[width=1.2\linewidth]{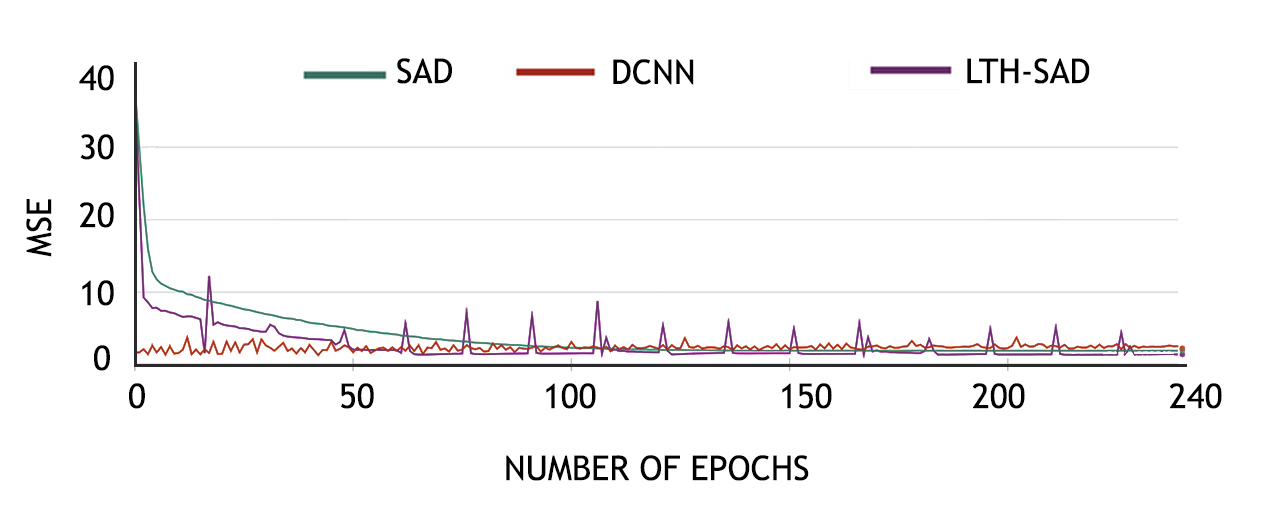}
        \caption{Comparison of MSE of LTH-SAD, SAD \& DCNN}
        \label{fig:topical_mse}
    \end{subfigure}
    \caption{Comparison of Textual Feature's MAE \& MSE}
    \label{fig:both_figures}
\end{figure}
\begin{figure}[ht]
    \centering
    \begin{subfigure}{0.45\textwidth}
        \centering
        \includegraphics[width=1.2\linewidth]{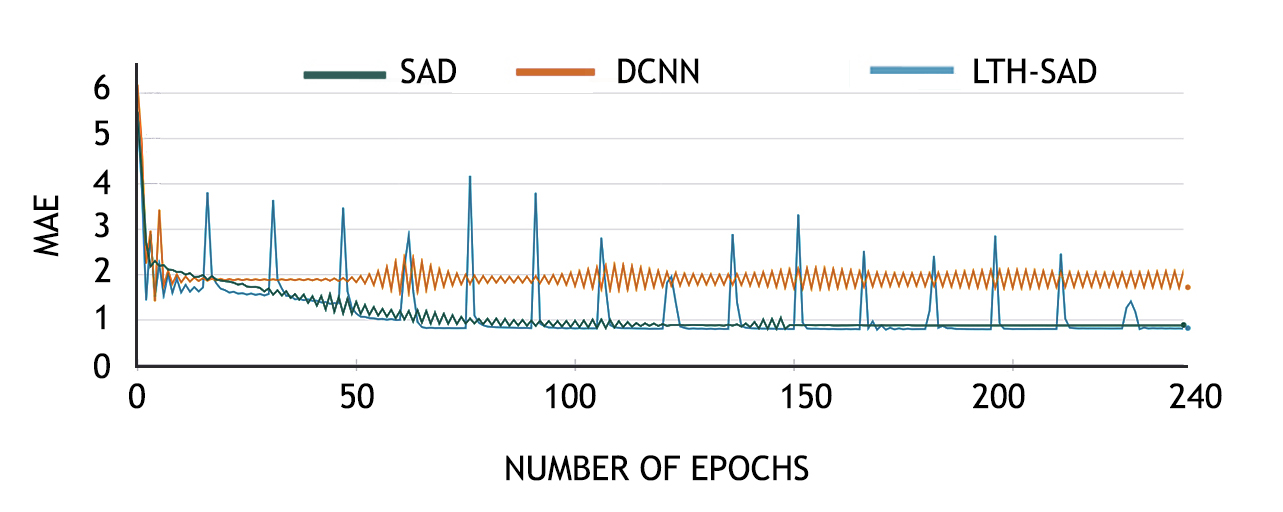}
        \caption{Comparison of MAE of LTH-SAD, SAD \&  DCNN}
        \label{fig:social_mae}
    \end{subfigure}
    \hfill
    \begin{subfigure}{0.45\textwidth}
        \centering
        \includegraphics[width=1.2\linewidth]{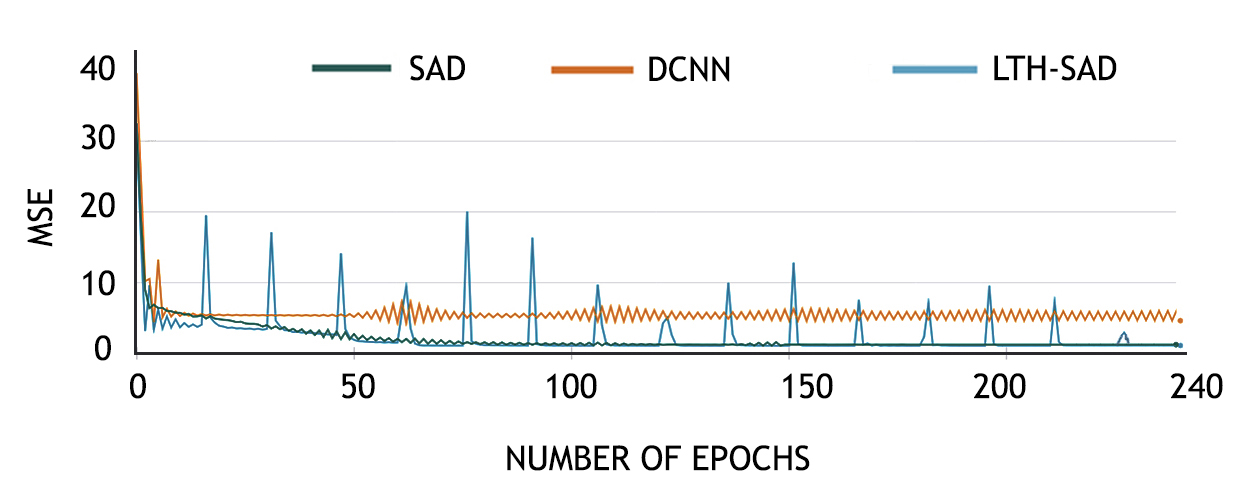}
        \caption{Comparison of MSE of LTH-SAD, SAD \& DCNN}
        \label{fig:social_mse}
    \end{subfigure}
    \caption{Comparison of Social Feature's MAE \& MSE}
    \label{fig:both_figures_social}
\end{figure}
However, we conceive that such outcome may be ascribed to a trade-off between complexity and generalization. Notably, attention mechanisms might not be as effective for certain types of data. To address this, incorporation of early stopping criteria and selecting features that does not introduce noise or complexities may improve the performance. 
The results have been presented in Figure~\ref{fig:both_figures} and Figure~\ref{fig:both_figures_social}. It is important to highlight that the periodic spikes in the LTH-SAD graph correspond to instances where the model is re-initialized using LTH (Lottery Ticket Hypothesis) after the pruning process. This approach ensures that the model begins learning anew each time, which leads to gradual but improved outcomes over time.

\begin{table}
    \small
    \centering
    \captionsetup{justification=centering} 
    \begin{tabular}{ |l|*{4}{>{\centering\arraybackslash}p{2cm}|} }
     \hline
     Methods & \multicolumn{2}{c|}{Categorical Features} & \multicolumn{2}{c|}{Numerical Features} \\
     \hline
     & MAE & MSE & MAE & MSE \\
     \hline
     DCNN & 0.8594 & 1.22 & \textbf{0.7455} & \textbf{0.98} \\
     \hline
     SAD & \textbf{0.8143} & \underline{1.183} & \underline{0.7714} & \underline{1.041} \\
     \hline
     LTH-SAD & \underline{0.8212} \textbf{(-8\%)} & \textbf{1.151} \textbf{(+2.7\%)} & 0.8607 \textbf{(-11.57\%)} & 1.255 \textbf{(-20.5\%)} \\
     \hline
     \multicolumn{5}{c}{} \\  
     \hline
     Methods & \multicolumn{2}{c|}{Social Features} & \multicolumn{2}{c|}{Textual Features} \\
     \hline
     & MAE & MSE & MAE & MSE \\
     \hline
     DCNN & 2.086 & 5.028 & 1.375 & 2.593 \\
     \hline
     SAD & \underline{0.88} & \underline{1.229} & \underline{1.163} & \underline{1.942} \\
     \hline
     LTH-SAD & \textbf{0.81} \textbf{(+7.9\%)} & \textbf{1.106} \textbf{(+10\%)}& \textbf{0.871} \textbf{(+25.1\%)}& \textbf{1.319} \textbf{(+32\%)} \\
     \hline
    \end{tabular}
    \caption{Performance Comparison of Features in Movie Recommender System}
    \label{Table4}
\end{table}

\textbf{Scalability:} We have used the GPU Power Consumption as a parameter of scalability where GPU Power Consumption takes into account the combined power consumed during training phase as well as inferencing phase. 
\begin{figure}[ht]
    \centering
    \includegraphics[width=0.8\textwidth]{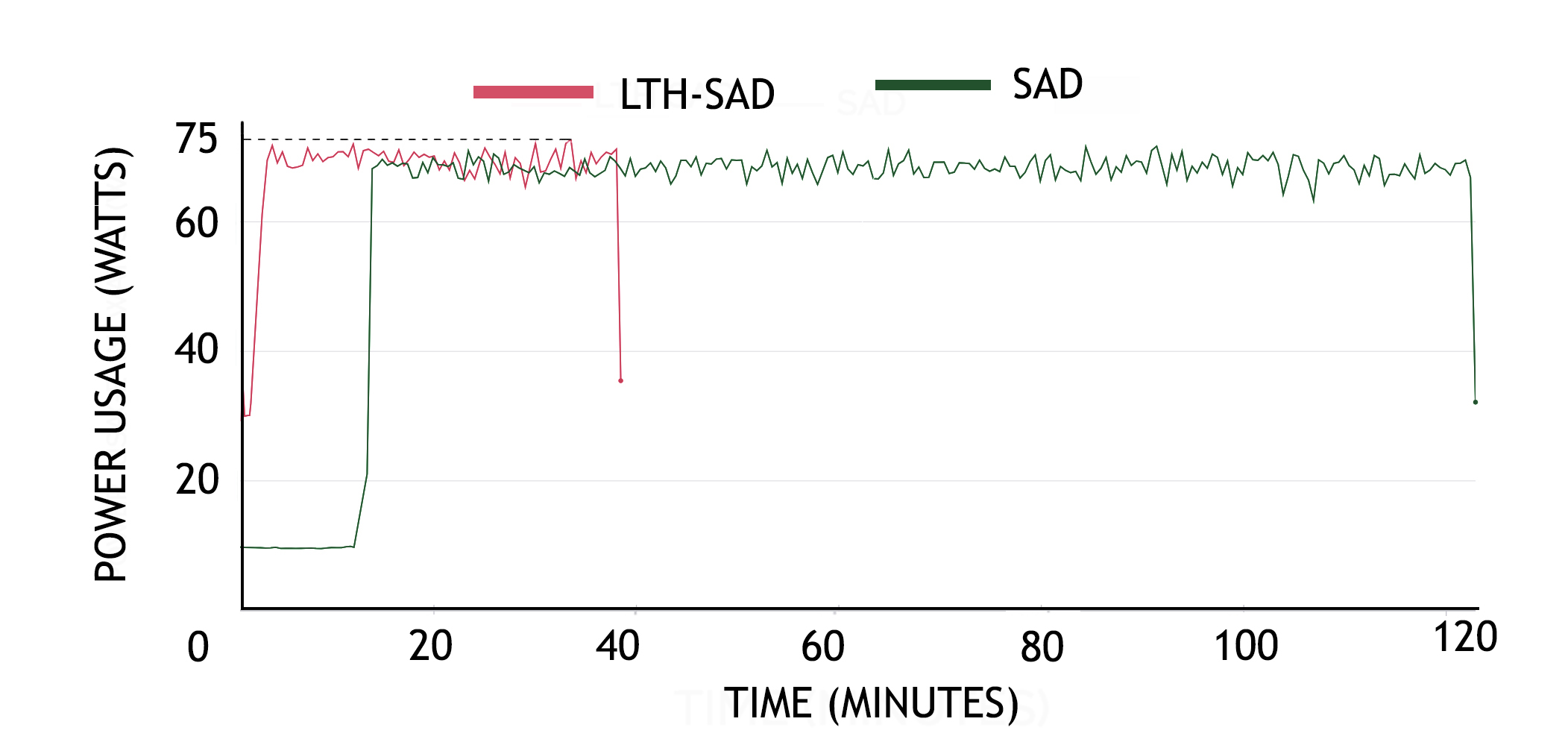} 
    \caption{GPU Power Consumption of LTH-SAD \& SAD}
    \label{fig:power}
\end{figure}
We have used movie dataset to experiment the power consumption by LTH-SAD and compared with baseline model SAD. Figure~\ref{fig:power} shows the power consumption with the models with categorical features. It is evident from the figure that SAD consumes 75W power for 120 minutes whereas LTH-SAD consumes 75W power for 40 minutes to yield a better MSE and MAE. This clearly shows LTH-SAD consumes 66.67\% less power with a 45\% reduced model size compared to SAD considering the dataset with categorical features. Similar results have been observed when other three features (numerical feature (9.09\% improvement), social feature (33.3\% improvement), and categorical feature (8.33\% improvement)) are considered. Overall, the outcomes of this study underscore the potential of our approach in facilitating the realization of efficient recommender systems suitable for real-world edge device implementations. 

%% file: P_7_Conclusion.tex
\section*{Conclusion and Discussions}

This research describes a unique method for improving the effectiveness and performance of deep neural network models by combining knowledge distillation with channel pruning. We allowed the student model to collect and distill the useful knowledge from the instructor model through the use of an attention-based framework and pruning, which resulted in enhanced generalisation and performance. To further minimise the size and complexity of the student model without compromising the accuracy and model size trade-off, we implemented three channel pruning techniques.
The efficacy of the models has been demonstrated by experiments on two diverse datasets. 
By combining knowledge distillation with structured pruning, our proposed models were able to produce small, effective models that are suitable for use in the contexts with limited resources. The results showed that, LTH-SAD achieved comparable accuracy (up to 32.08\% , 25.10\% improvement in MSE and MAE, respectively) with reduced GPU power consumption (up to 66.67\%) with 45\% reduction in model size. The outcomes confirm that compared to SAD \citep{ji2021show}, LTH-SAD performs better with the textual and social features. 
Our approaches further attained favourable outcome in the context of recommender systems for rating prediction to identify key qualities or connections between the input data (such as reviews, user preferences, or movie attributes) and the projected ratings. Future work might also study the use of our method in numerous domains and real-world situations, as well as other optimisation strategies. 